\documentclass[preprints,article,accept,oneauthor,10pt,a4paper]{mdpi}

\firstpage{1}
\pubvolume{1}
\issuenum{1}
\articlenumber{0}
\pubyear{2022}
\copyrightyear{2022}
\externaleditor{Academic Editor: Mariusz P. Dąbrowski, Adam Balcerzak, Vincenzo Salzano and Gonzalo J. Olmo}
\history{Received: 31 January 2022; Accepted: 05 April 2022; Published: 08 April 2022}

\pdfoutput=1

\usepackage{amsmath}
\usepackage{amssymb}
\usepackage{hyperref}
\Title{Ultraviolet Finiteness or Asymptotic Safety in~Higher Derivative Gravitational Theories}
\Author{{Les\l{}aw Rachwa\l{}} $^{1,\,{{\dagger}}}$\orcidA{}}

\AuthorNames{Leslaw Rachwal}

\address{%
$^{1}$ \quad Departamento de F\'{i}sica, Instituto de Ci\^{e}ncias Exatas, Universidade Federal de Juiz de Fora, \newline Juiz de Fora 33036-900, MG, Brazil; grzerach@gmail.com\\
$^{{\dagger}}$ \quad  {Speaker at ALTECOSMOFUN'21 (Alternative Gravities and Fundamental Cosmology) Conference}
}

\abstract{We present and discuss well known conditions for ultraviolet finiteness and asymptotic safety. The requirements for complete absence of ultraviolet divergences in quantum field theories and existence of a non-trivial fixed point for renormalization group flow in the ultraviolet regime are compared based on the example of a six-derivative quantum gravitational theory in $d=4$ spacetime dimensions. In this model, it is possible for the first time to have fully UV-finite quantum theory without adding matter or special symmetry, but by inclusion of additional terms cubic in curvatures. We comment on similarities and some apparent differences between the two approaches, but we show that they are both compatible to each other. Finally, we motivate the claim that actually asymptotic safety needs UV-finite models for providing explicit form of the ultraviolet limit of Wilsonian effective actions describing special situations at fixed points.}


\begin{document}

\section{Introduction}
\label{s1}

The behaviour of quantum gravity (QG) models at very high energies, so in the ultraviolet (UV) regime, was always of big interest for research in theoretical gravity.
Quantum considerations in the perturbative framework force us to use modified gravity models as viable and consistent models for QG dynamics. This is because of the well known fact of quantum non-renormalizability of the theory based on the Einstein--Hilbert gravitational action. In this note we want to compare the methodology and conditions between ultraviolet-finite and asymptotically safe theories. We will consider them on a particular example of higher derivative gravities in $d=4$ spacetime dimensions. To be definite, the model with six derivatives in the gravitational dynamics will be for these purposes studied here. After the considerate analysis, we decided that the $6$-derivative gravitational theory, being  in the general class of higher derivative local gravitational models, better shows the differences between the two approaches for UV-completion of gravitational theories. Namely, we believe it does this better than  $4$-derivative Stelle quadratic theory, which in $d=4$ is simply perturbatively renormalizable model \cite{Stelle:1976gc}. In a pure gravity case, this last model is not completely without perturbative UV divergences (is not UV-finite), but it can reach a non-trivial fixed point of the renormalization group (RG) flow in the UV regime (can give rise to the asymptotically safe model as found in~\cite{Codello:2006in}). The longer discussion of this issue is presented in Section \ref{s2.1}, where we point out that the conformal window is not possible for pure Stelle gravity. However, the case of pure six-derivative gravity offers for us such a possibility, probably for the first time. Pure refers to a gravitational theory which is a quantum theory of only metric degrees of freedom, without addition of matter fields or additional symmetry (like in supergravity cases). Such a modified gravity model with six derivatives in its defining classical action, first studied on the quantum level in \cite{Rachwal:2021bgb}, is a good and interesting theoretical laboratory for investigations of various perturbative and also non-perturbative RG flows.

In the framework of quantum field theories (QFT) of fundamental interactions, one of the biggest theoretical problems is the proper form of UV-completion of theories which work well at some low energy regimes. One could simply view them as effective theories valid in these energy ranges. As such, they must work well there. The interesting conceptual question is what happens with these theories when the energy scale is increased. There are various possible options here---one positive is that the model in question can be renormalizable (perturbatively or non-perturbatively). However, this is not the end of the story in the UV regime since even some renormalizable theories may meet some problems on the way towards the UV regime. A possible danger is for example, the presence of the Landau pole for running coupling parameters. Then, such models are not UV-complete. A first possible resolution is to require that the models are not only renormalizable but that they are also asymptotically free (AF) in the UV regime and that the couplings tend to vanishing values in the high energy limits. This scenario is realized, for example in the UV-consistent QCD---theory of strong interactions. We remark however, that in such a model the UV divergences are still present but that they are fully under control and hence are not anymore dangerous and that the couplings do not run to divergent values at any finite (or at infinite) energy scales.

The renormalizability brings the additional control over divergences that one encounters by doing higher order (loop) computations in the QFT framework within the perturbative calculus very frequently used for considerations of quantum corrections. Unfortunately, this is not the case for Einsteinian gravity in $d=4$. In the latter theory the control over perturbative infinities is lost when the situation is considered off-shell, at the two-loop  or higher order of accuracy, or when the matter fields are added to the quantum~system.

One of the way out for the problem of infinities in QG, is to consider higher-derivative theories in $d=4$ starting first with quadratic gravity of Stelle.  Then, the renormalizability is regained, but the price is the higher-derivative nature of classical and quantum dynamics with various instabilities. Another option as proposed by Weinberg in 1979 \cite{Weinberg:1980gg} is to solve the problem with perturbative UV divergences by demanding an existence of a non-trivial fixed point (FP) \cite{Smolin:1981rm} of RG flow in the UV regime. This last scenario  is inherently non-perturbative in values of coupling parameters since at the FP the running couplings must attain some finite non-zero values. In this way the situation differs from the trivial or Gaussian FP, where the FP values are vanishing.  We also note that in AS the problem of higher derivative instabilities is probably solved by the arguments which were presented in \cite{Platania:2020knd}, which show that
the Boulware--Deser ghosts are most probably fictitious particles. 
 Contrary to the solution proposed by Weinberg of the problem of UV divergences, usually in the perturbative framework, another approach is proposed. Namely one seeks to  find a theory so special and tuned that all UV divergences disappear that there are no  issues with them and all the perturbative loop integrals are perfectly finite regarding the UV limits of integration.

The above two approaches for definitely solving the problem with UV infinities have a lot in common. There are however some subtle but important differences that we want to discuss and oppose in this article. We believe that the analysis of such dissimilarities will be beneficial for bringing together the research directions and for finding a unified description of QFT of QG which is without any problem at arbitrary high energy scales. For this goal, we will present the results of the analysis for the promising model of six-derivative gravitational theories, which can be made completely without infinities in the UV regime.

In full generality, a consistent model of QFT of fundamental interactions, like in the case of QG for metric fluctuations $h_{\mu\nu}$ for the quantum graviton fields, must be defined as an interpolating theory between two FPs of RG (or in the special case, like of QED, we must invoke some threshold phenomena in IR to decouple massive charged modes). This relates to the necessity of having the possibility to define the interpolating $c$-function as this was emphasized in \cite{Anselmi:1997am,Anselmi:1997ys} for any well-defined QFT model. One of the two FPs may be in the infrared (IR) regime and we will not discuss here the infrared safety notion related to it. The second FP must be in the UV regime and its presence there is required for UV-completion of the QFT model.

It is well known that QFT valid at all
energy scales (without any other form of UV-completion like by string theory or by some quantum theory of some other non-local objects) should be
defined as an interpolating QFT between two FPs---between IR and UV regimes. These two endpoints FP of RG are
musts, otherwise for finite or infinitely large (or infinitely small) energies the RG flow in the theory reaches a dangerous Landau pole, and this means that the
description of the quantum system in terms of continuous field theory of point-like particles fails, or at least that the perturbativity fails and one has
to resort to study some non-perturbative effects here. For the question of UV consistency the IR problems do not matter and the IR FP is unnecessary. The safe situation for the IR regime of the theory can be obtained, for example, if threshold phenomena decouple all interacting modes (by giving them mass) and only free (Abelian) massless modes remain, then there is no need for IR FP of interactions since there are no IR interactions. There exist possible UV-completions by a quantum theory of something else and these are not relativistic fields and we will not discuss them here. However, here and in the QFT approach to QG, the field-theoretical description of the quantum system is assumed to be valid at all energy scales. Hence QFT of fundamental interactions must be defined with the sense at all energy scales and the two FPs must exist (some of them may be of course trivial Gaussian FP describing a free non-interacting theory).

Next, these two FPs may be found in QFT of completely different
sets of active degrees of freedom from the point of view of UV and IR FPs, respectively. So we must allow for the possibility of transmutation and of
changing the nature of degrees of freedom of the theory in the course of RG flow. Still, we require them to be quantum fields with point-like nature of excitations around the vacuum. We should also allow for the impact of the important effects due to threshold phenomena when some degrees of freedom acquire mass, they decouple, and are effectively integrated out on the level of functional integral. In such a case, they are absent from the spectrum at lower energies, towards the IR limit. The
vacuum for these theories (both in the UV and IR regimes) is always the same since it has to be defined non-perturbatively and exactly as
the stable state with minimal value of energy. This definition is independent of the energy of quantum fluctuations, regardless whether
they are at UV FP (infinitely large energies) or at the IR FP (infinitely small energies). All this is needed for the consistent description of the scheme
of QFT valid at all energies and with the RG flow in it which interpolates between the two FPs.

In the next two subsections we briefly discuss the main ideas behind the theoretical notions of UV-finiteness in quantum field theories and also behind the asymptotic safety (AS). The second section will be devoted for a short presentation of the six-derivative gravitational theories in $d=4$. We will also quote the main results for exact beta functions of perturbative couplings there. In the main section of this paper, we analyze the similarities and differences between AS and UV-finite scenarios based on the example six-derivative theory. Finally we draw our conclusions and comment on a possibility of unified framework and how UV-finiteness should be incorporated in the UV safety scenario.

\subsection{UV-Finiteness}
\label{s1.1}

In the first case, one typically relies heavily on the perturbative approach. It is not so surprising since the first place where one meets dangerous UV divergences is during the expansion of quantum predictions of general QFTs in small coupling parameters. The ultraviolet divergences are the most ubiquitous theoretical phenomena appearing in almost any QFT models. Contrarily, if the QFT model is very special, then these infinities may be absent. Such situations typically call for an enhancement of symmetries to full quantum conformal symmetry. However, here we want to speak about perturbative UV divergences. If they are absent in all orders of perturbative calculus, then we can say that the theory is UV-finite. Since, for example, at the one-loop level, these UV divergences are related to the perturbative beta functions of the couplings of the theory, then an equivalent reformulation of the conditions for UV-finiteness reads
\begin{equation}
 \beta_i=0,
 \label{condition1}
\end{equation}
where the index $i$ counts all the couplings $g_i$  of the theory. For this the general theory we write in the form defined by the action $S=\sum_i g_i {\cal O}_i$, where ${\cal O}_i$ lists all classical terms built out of fields and derivatives of the model.

The condition in (\ref{condition1}) is not typically satisfied for any value of the perturbative couplings $g_i$  of the model. The situation must be very fine-tuned to have $\beta_i=0$ to all loop orders (and also non-perturbatively). This requirement must also hold at any energy scale since the running energy scale $E$ is never an explicit argument of the equation in (\ref{condition1}). In the language of RG flows such a special theory sits at the FP of RG all the RG time and never leaves it. This is obvious since RG running phenomena arise from the ability to absorb all UV-divergent quantities in bare non-observable parameters of the model. In the case of UV-finite theories, there is no RG running and the theory is scale-invariant on the full quantum level. Likely this last symmetry can be promoted to the full quantum conformal~invariance.

There are few known examples of theories that they are completely UV-finite in the standard relativistic QFT framework. Enhanced symmetry helps in finding such models, but this is not a necessary ingredient. We can mention here an example of ${\cal N}=4$ super-Yang-Mills theory \cite{Namazie:1982br} and also of ${\cal N}=4$ supergravity due to Fradkin and Tseytlin \cite{Fradkin:1985am}. These models require the highest possible for consistency level of supersymmetry in their formulations. However,  in other gauge-scalar-Yukawa models as found by Litim and Sannino in~\cite{Codello:2016muj,Litim:2014uca}, the UV-finiteness is with a less degree of symmetry.

Generally, it is very difficult to find QFT models completely without any perturbative UV divergences at any loop orders and such UV-finite theories are very special. Of course, they do not show any problem in the UV regime and they are definitely UV-complete quantum models of field theory. Strictly speaking UV-finiteness should also hold non-perturbatively and one should avoid in such models also possible UV divergences that one could find in some non-perturbative approaches to QFT. We envisage therefore the possibility that infinities and divergences can be also non-perturbative in some schemes. The condition for complete absence of UV divergences is typically a non-linear condition in all couplings of the theory and first it is difficult to find one. Secondly it is even more difficult to solve it for couplings and in this way define completely the position of the FP where the RG flow effectively stops.

In some matter theories (without gravity) UV-finiteness may be viewed as some additional condition on the quantum dynamics. It gives rise  to enhanced conformal symmetry on the quantum level, but in matter models as we have seen for QCD this is not essential. The problem with UV-completion still may be solved even if the UV divergences are present even in the asymptotically very high energies. This only means that one has to perform renormalization of these divergences also at infinite energies.

However, in the gravitational setup UV-finiteness of some QG theory is a very desirable feature since only with it consequently promoted to full conformal invariance in the general-relativistic (GR) setup, we can deal with the problem of singularities of the classical gravitational field. As emphasized in \cite{Bambi:2016wdn,Modesto:2016max,Rachwal:2018gwu} full conformal symmetry on the quantum level is a key ingredient for the successful resolution of singularities. One of the prerequisite for this is that the quantum theory should be completely without divergences and conformal symmetry should not be broken at very high energies. Of course, for the issue of singularities we need only to study the UV regime of the theory. The important question here is how quantum corrections help or destabilize the solution of the problem with these pathologies in classical theory. An example of the theory which solves these issues on the classical level is a conformal gravity (Weyl gravity) in $d=4$ spacetime dimensions. For the stabilization of these resolutions one has to ensure that quantum effects do not destroy this conformal symmetry of the classical theory, or in other words, that conformal symmetry is also preserved in the quantum theory. This is achieved when there is UV-finiteness and there are no divergences and the quantum effective action also enjoys the important feature of being conformally symmetric. Hence UV-finiteness is a must for consistent QG model if it has to deal also with spacetime singularities.

On the other side, one has to recover the Einsteinian GR in the IR limit as it was explained above. However, the problem of
infinities is in the opposite UV regime. The situation at the highest energies should be very symmetric
and one can view that the IR phase is a spontaneously broken phase, where the conformal symmetry is not
present in the vacuum state of the theory. From the UV-consistent theory we require the complete
cancellation of infinities and reaching the UV conformal phase. Moreover,
this conformal phase does not extend for all energies, especially for the low energies (or even Planckian
energies). In the IR the breaking of quantum conformality is expected \cite{Jizba:2020hre} and is welcomed there to provide a
consistent description of IR low-energy gravitational phenomena, like gravitational scattering
amplitudes, cosmological perturbations \cite{Jizba:2014taa}, gravitational threshold phenomena \cite{Jizba:2019oaf}, etc.

\subsection{Asymptotic Safety}
\label{s1.2}

Asymptotic safety (AS) is an approach that in turn generalizes the notion of asymptotic freedom for all running coupling parameters \cite{Niedermaier:2006wt}. It describes this both in the perturbative as well as in non-perturbative attitude towards quantum physics.

The idea of Asymptotic Safety scenarios came from Weinberg and it was proposed in 1977 \cite{Weinberg:1980gg}. This was in close analogue to asymptotic freedom (AF), which makes
strong interactions not so strong in the UV regime due to non-trivial RG flow of Yang--Mills coupling. The idea of asymptotic safety was
that many problems of Einstein quantum gravity are also solved in a similar way due to non-trivial RG flow of coupling parameters in the
gravitational theory. One of the problem is the famous non-renormalizability of Einstein--Hilbert (E-H) gravity (with or without cosmological
constant) and in $d=4$ spacetime dimensions, off-shell, at one-loop level and coupled to some mater species, or also in pure gravity at the
second loop level as proved by explicit computation of UV divergences by Goroff and Sagnotti \cite{Goroff:1985sz,Goroff:1985th}. The other issues concern the unitarity
bound on the scattering amplitudes of graviton-graviton interactions, which is apparently broken in the theory based just on E-H action at
energies of the scattering gravitons being of the order of the Planck scale. Although it is well known that the E-H gravity is unitary
two-derivative theory these two problems deal really with the UV regime of the quantum gravitational theory, where the simple E-H theory
is untenable on the full quantum level.

The idea of Weinberg to solve the non-renormalizability issue and the growth without bound of the quantum scattering amplitudes was to
invoke the situation that in the UV regime a non-trivial fixed point (FP) is met of the RG flow of running coupling parameters in the
gravitational action, which is typically taken as different from the one given by the  E-H truncation action.
The values of the couplings are non-zero and can be also not very small, so the existence of such a FP is truly a non-perturbative issue to
settle which one probably needs to use methods of non-perturbative field theory and RG flows there. One still can hope that such a
non-Gaussian FP  can be reached by methods of perturbation theory in some various small deformation parameters (not necessarily like
the coupling)---they could be  dimension of spacetime away from $d=4$ or similar ones. The existence of non-trivial FP  apparently solves
the problem of non-renormalizability since now in the UV regime the theory is without divergences, without RG flows, since the FP is met. 
 This  is also often called as non-perturbative renormalizability. Although for example, the issue of
non-perturbative form of UV divergences of the theory and whether they are present here or not is only rarely considered.  In turn, there may exist also non-perturbative divergences of the QFT models and the issue of their reabsorption in some models of field theory is well posed too.  Speaking maybe too conservatively, one could say that in this last sense one has non-perturbative renormalizability, if even non-perturbative divergences are possible to be absorbed in the counterterms of the form identical to the terms present in the original action of the model. After all, the notion of the effective action of any QFT model is not restricted only to the perturbative calculus, and also its divergent part has the meaning independently of using it. Then the issue of renormalizability of these non-perturbative UV-divergences is sensible too and this would constitute the core part of the conservative definition. Of course, then the UV-completion by non-perturbative UV FP of RG flow for all couplings is conceptually something else and this requires the departure from the original conservative definition of what was renormalizability. To support more modern point of view we can remark that Wilson and others \cite{Wilson:1973jj} have formally shown that the existence of a non-trivial UV FP (with non-perturbative values of couplings) guarantees  renormalizability of the theory in the first sense (possibility to absorb all perturbative UV-divergences of the model in question computed via loop integrals in counterterms of the original action). Moreover, the number of parameters needed to reabsorb such divergences equals the number of relevant directions of the RG flow near that UV FP.

One shall not forget the reasons why we have RG flow in the QFT at the first place. Historically first classical models of field theories were with constant coupling parameters. Then one could have thought  of models with spacetime-dependent coupling parameters as unnecessary complication since they would typically break homogeneity of the spacetime. Originally coupling parameters in field theory were constants. The great discovery and also the great puzzle of interacting QFT  in the perturbative framework came with infinities. They were unwanted, but theoreticians had to learn how to live with them and to deal with them in such a way to bring the predictions back to finite values. This was finally achieved in the advent of renormalization of infinities in QFT. The divergences were reabsorbed in counterterms of the theory. This was possible in any renormalizable QFT model. However, there was also a price to pay for this possibility---the couplings could not be anymore constant, they would have to show dependence on the arbitrary renormalization energy scale $\mu$. In this way they could be viewed as dependent on the fifth external dimension being the energy. However, this is fully consistent with relativistic symmetries of the underlying spacetime, if the extended 5-dimensional background spacetime is maximally symmetric and negatively curved.  The dependence on the energy as the dependence on additional dimensions can be further developed in the framework of AdS/CFT correspondence and its generalizations. Coming back to the perturbative framework, the reason for RG flow is traced back to the presence of UV divergences. If the latter were not present, then we would have only finite renormalization of couplings and not an ensuing RG flow. Hence all the story with perturbative RG flows started because of infinities. One can very carefully isolate them from the functional of the quantum effective action of the theory. There they could be both of the perturbative and also non-perturbative character. Hence the notion of renormalizability and its further generalizations should  primarily deal with these divergences and the possibility of their reabsorption in the original terms of the theory named here as counterterms. Additionally, this story of UV divergences can be finally closed too, if the theory is completely UV-finite. Therefore UV-finiteness is a very natural generalizations to the notion of (perturbative) renormalizability.

It is  however a fact that at the non-trivial (non-Gaussian FP), the theory by definition does not show any RG effects. The issue with
non-renormalizability is more tricky here, since one cannot say then whether the perturbative or non-perturbative UV divergences are still there or not. As a matter of fact about divergences presence one derives by looking at the asymptotic behaviour of various integrals, so then a special limiting procedure must be performed to accomplish this task. Instead, the situation with an UV FP of RG is that one is sitting at the FP already and for all the time,  and then one cannot change the energy scale and cannot perform any asymptotic limit to see the behaviour when the FP is reached. It is true that what is really important is not that whether the FP exists or not but rather how it reached there in the UV regime. We can concentrate just on the leading asymptotics with the most relevant behaviour. One could derive whether the perturbative (or also non-perturbative) UV divergences are there so whether the problem with non-renormalizability is ameliorated or not.  For example, from the paper by Kogut and Wilson~\cite{Wilson:1973jj}, it was established that the FP with a finite number of IR-relevant directions gives rise to renormalizability in the perturbative scheme.

As mentioned above, in the AS programme for UV-completion of some interacting QFT models, one is more interested in the way how the FP is reached in the UV than just the location of this FP in the parameter space (moreover, the last one is not invariant under arbitrary coupling redefinitions, hence it does not possess any observable physical meaning). To find a fixed point of RG, one studies the solution in the parameter space of the equation $\beta_i = \frac{dg_i(t)}{dt} = 0$, where $t$  is the variable of the logarithmic RG time. To quantify the way FP is reached, one must study higher derivatives of the flow of the couplings $g_i(t)$, for example the second derivatives $\beta'_i = \frac{d^2g_i(t)}{dt^2}$ and evaluate them at the FP, so with the condition that $\beta_i=0$. However, from the theoretical point of view much better quantities are the following derivatives put in the matrix form $M_{ij} = \frac{\partial\beta_i}{\partial g_j(t)}$, which do not show now any explicit $t$-dependence and they are autonomous in the system of all running  couplings $g_i=g_i(t)$. The eigenvalues of such a matrix (so called stability matrix) of derivatives are related to the so called critical exponents of the theory at the FP. This issue is studied extensively in all FRG papers, since these critical exponents are thought to be observable physical quantities as we know, for example, from the theory of critical phenomena applied to the condensed matter physics. Close to the FP one has non-trivial scaling of couplings in a form of various power laws that are parametrized by these critical exponents. The computation of these critical exponents is one of the main task of the theoretical description of the situation at the FP. 

Moreover, as it was explicit in the original Weinberg's construction besides the existence of a non-trivial FP in UV to make the theory safe
(from the problems of the UV regime), one must also ensure that the dimensionality of the critical surface on which such a FP lays is finite.
Such a surface is typically parametrized by the relevant directions from the point of view of the FP in question, which is located in the UV
regime. Then this classification of the ir/relevant directions is sensitive which FP this is -- in the UV or in the IR we speak about. To each
relevant direction correspond an operator in the action of a theory that when it is included it drives the RG flow away from the FP. Simply
unmodified theory is at the FP, while the deformation by adding some operator breaks the scale invariance and the RG flow starts taking
us away from the FP when the energy scale is lowered compared to the formally infinite abstract theoretical energy scale of the UV FP.
The deformation action is of the general form
\begin{equation}
S_{\rm def} = \int d^4x\sqrt{|g|}g_i{\cal O}_i,
\end{equation}
where we summed over all deformation parameters and their coupling parameters $g_i$. Each of this coupling should reach a finite FP
value in the UV, and away from the FP we should see its non-trivial RG running.

Additionally, if the operators are (classically) marginal from the point of
view of the non-trivial UV FP, then the running coupling at least in the vicinity of such FP should be logarithmic with the energy scale, but
the value it attains in the limit of infinite energies should be finite and typically non-zero (for a non-Gaussian FP). There are known
examples of the studies how such non-trivial FPs can be reached when going towards the UV regime, with explicit both perturbative (at the
one-loop level) and non-perturbative forms of the running of the coupling parameters.

The second requirement from Weinberg about finite
number of relevant directions is very important for the predictability of the theory since the theory with FP existing but with infinitely many
ways how it can be reached is not useful physically since this vast infinite freedom of choices for infinitely many relevant couplings does
not constrain at all the physics at some finite intermediate energy scale, which is neither in the UV, nor in the IR. Of course, we could also add
an arbitrary (possibly even infinite) number of irrelevant deformations from the point of view of UV FP with some irrelevant coupling parameters. The scaling of such irrelevant couplings would be $g\sim g^*+c k^\alpha+O(k^{\alpha-1})$, with $\alpha>0$. For relevant couplings (in my terminology) one would have that $\alpha<0$.  We remark that since relevant couplings from the UV FP perspective  run to fixed values $g^*$, when $k\to+\infty$, so towards the UV FP, they do not change or modify anything regarding the question of how the UV FP is reached. This is why they are irrelevant from the point of view of the UV physics. (However, they can be relevant from the point of view of IR, where also their scaling exponent $\alpha$ may change to something else like $\alpha'$ possibly being or positive or negative.) As for the issue with irrelevant couplings with $\alpha>0$ the situation with them must be definitely under full control and they are not unconstrained as relevant couplings were at the UV FP. For $\alpha>0$, one has that the running in the UV leads to a blow-up $g\to+\infty$, unless $c=0$. In order to avoid this and reach the FP, one has to lie on the UV critical surface, so the initial condition is such that the constant $c=0$ for all irrelevant deformations (in terminology with $\alpha>0$) and thus at the level of the effective action one would have all such  operators multiplying the corresponding (generally non-zero) fixed-point values of the couplings $g=g^*$. This also means that all possible operators consistent with symmetries and particle content of the theory must be kept in the action (unless by chance we have that $g^*=0$). However, in the case of six-derivative gravitational theory we also add here a requirement from effective field theory that the number of derivatives in them is not bigger than 6.

Finally, we could also add some marginal operators with marginal logarithmic running
RG flow of couplings, but here as usual with marginal deformations, the situation needs further attention (and for example going to the
higher order in the perturbation theory) to finally pinpoint the issue whether they are marginally relevant or irrelevant operators due to
higher order corrections. They then could fall into one of the two categories described above.

The conditions for asymptotic safety (AS) should solve the problems with perturbative non-renormalizability of some theories (like
quantum gravity) and also should tame the growth of some scattering amplitudes beyond the unitarity bound and also up to infinity for
infinite energies. The operational definitions of couplings due to Weinberg is to read them from some observed physical process, when it is sure
that they are physical observables and they are well defined and measurable, at least in principle. Then the physical requirement of AS
means that the physical scattering amplitudes or cross sections tend to some finite values for the energy of scattering particles tending to
infinity. The theory is then indeed safe regarding the possible infinities present in the observables of the theory. However, this very
operational scheme is very difficult to realize in practice for theoretical computations and for finding evidence for the existence or not of
the fixed point and for the characteristic of the critical surface, on which it lays. One rather uses the theoretical and abstract
scheme of computation with abstract energy scale and computes correlations in terms of abstract, non-physical coupling parameters of the
theory.  For example, in AS approach to the UV-completion, this is done right away using the vertex expansion \cite{Pawlowski:2020qer} and references therein.  
Only later, in standard perturbation calculus approach, one has to physically dress all the quantities computed theoretically like Green functions (on-shell), scattering
amplitudes \cite{Christiansen:2017cxa,Knorr:2019atm,Denz:2016qks,scattering} and couplings. The usage of this scheme in principle leads to some other important questions. We have to keep in mind that AS makes the situation under
control in the deep UV regime, asymptotically at the point where the energies are formally infinite. However, in practical aspects of quantum
gravitational physics we will be more interested in situation at some intermediate energies, where the energies are of the order of Planck
energy or beyond (trans-Planckian energies) and make the situation safe there regarding the scattering amplitudes.  The RG flow from the UV to the IR has also been studied in many papers in the framework of AS and in functional renormalization group (FRG), e.g., in~\cite{Gubitosi:2019ncx} (pure gravity and in the quadratic approximation) and also in~\cite{Eichhorn:2019ybe}  (coupled gravity-matter systems).

The main conditions for AS are formulated in such a way that they touch upon the issue of existence of a non-trivial fixed point (FP) in the UV regime, which must also lay on a finite-dimensional critical surface parameterized locally in the vicinity of this FP by a finite number of relevant directions. All these relevant directions correspond to the deformations that could be added to the original action of the theory at the UV FP, so to the UV action. Then one violates the scale-invariance present in the FP and causes a non-trivial RG flow down with energies. The relevant operators with respect this UV FP parametrize locally the tangent space to the critical manifold where the touch point is precisely the FP in question. Besides these conditions imposed in the UV regime of the theory, for the physical consistency and phenomenological viability  the RG flow in the whole parameter space down with energies must also possess some further special features when it is considered on the phase diagram. We will not consider them in more details here. We just want to remark that the scenario for AS or UV-finiteness regards really the situation at infinite energies, where the theoretical problems are the most severe. However, one knows that to compare with QG phenomenology one must really study the RG flow phenomena at physical energy scales, like around the Planck scale. The conditions in the UV regime should leave some imprint and when one runs down with energies, then for example the effective action functional should be constrained because of this very particular asymptotics in the UV regime.

\subsubsection*{AS for Dimensionless Couplings and Dimensional Transmutation for Dimensionful~Ones}
\label{as}

Finally, when looking for pieces of  evidence for AS  scenario in the UV regime, we need to discuss the beta functions $\beta_i$ of which couplings $g_i$ we will seriously consider for finding a non-trivial UV FP of RG. In the perturbative calculus, it is well known that only logarithmic divergences in the formal UV-cutoff (as put in momentum space) are universal and unambiguous in any models of QFT. The infamous power-law like UV divergences (cf. their description in \cite{Anselmi_reno}) are not in such a class since they can be easily reabsorbed and they do not require and they do not cause any RG flow. (After all, we remind the reader that the true reason for the RG flow of coupling parameters, which were originally thought of to be always constant in the classical physics, is that there are non-absorbable in the field redefinition logarithmic UV divergences that need counterterms and renormalization and hence also introduction of bare and running coupling parameters $g_i(t)$. In the last expression we denote by $t$  the logarithmic RG-time which is defined as $t=\ln({k}/{k_0})$  with some reference energy scale $k_0$ and with the running energy scale $k$.) The power-law UV divergences are ambiguous and for example they depend on the scheme of regularization used to find them, hence the information contained in them can never be physical and will never have any influence on true observables of theoretical models in QFT.

Due to the above reasons, we only study the AS conditions  for couplings whose RG flow is determined in the perturbative approach from logarithmic UV divergences. If the theory is only with higher derivatives and does not contain in their original defining action any subleading terms with the smaller number of derivatives, then the couplings with the above description are simply dimensionless gravitational constants. For example, in $d=4$ these are couplings in front of GR-covariant terms of dimension four, so in front of $R^2$ and $C^2$ terms, where $R$ denotes the Ricci scalar and $C$ the Weyl tensor. Their RG flow is read from logarithmic UV divergences. We will only concentrate on them in the QG models without subleading terms. The six-derivative gravitational theory that we will present results of in the next section is precisely of this type. The running of other prominent gravitational couplings (like of the Newton's constant $G_N$ or of the cosmological constant $\lambda$) in the minimal six-derivative gravitational models is absent, while when one adds also some subleading terms in the number of derivatives, like $R^2$, $C^2$ or the E-H $R$ term, then the RG flow for them is back and it should be then read exclusively from logarithmic UV divergences now present also in such an extended model. The computations of this type were first presented in \cite{Asorey:1996hz,Modesto:2017hzl}. 

A corresponding FRG study of the system with six derivatives has been done recently in \cite{Knorr:2021lll} to the one-loop perturbative level, extending at the same time the original work on Stelle gravity from \cite{Julve:1978xn,Fradkin:1981iu,Avramidi:1985ki} and later from \cite{Codello:2006in} to the case of $d=6$ spacetime dimensions. In this new setup much more non-trivial FPs were found. This can be viewed also an extension of our work with six-derivative quantum gravitational dynamics to the case of six dimensions, where the latter theory is a minimal renormalizable gravitational model. However, as it is obvious from counting of constraints $\beta_i=0$ for all 10 couplings of such a model in $d=6$ and comparing this with the number of active couplings of the theory on which one-loop divergences are dependent, such a minimal model cannot be UV-finite, although of course there is much more UV FPs found in the perturbative AS framework. The reason is the presence of the topological term (an analogue of the Gauss-Bonnet/Euler term from four dimension) which does not give any impact for one-loop UV divergences. However the UV divergence proportional to this term and the corresponding  beta function for this coupling has to be made vanish in fully UV-finite model. Hence we see that we have by one too many equations than the number of independent couplings (the value of the topological coupling is arbitrary---it is completely unconstrained from the conditions of perturbative UV-finiteness) and such an algebraic system is generically without solutions. One can resort here to getting partial UV-finiteness only in some subsectors of couplings, similarly like this was done in the case of quadratic gravity in \cite{Codello:2006in}. It is true that in the framework of the paper \cite{Knorr:2021lll}, the cubic terms like ${\cal R}^3$ are present; however, they do not play the role of cubic killers since couplings in front of them in the minimal renormalizable model (which must have six derivatives in $d=6$) do exhibit RG running (perturbative even at the one-loop level). This in clear contradiction to the six-derivative model as considered here, where the couplings in front of cubic killers operators \cite{Rachwal:2021bgb} do not run. Therefore, if one would like to make the presented model in six dimensions completely UV-finite, then one should first promote the theory to super-renormalizable by including 8-derivative or even higher 10-derivative kinetic terms in the action. In the next step, one could use quartic killers (of the type ${\cal R}^4$) to kill all the beta functions, although then the algebraic system of equations is non-linear. If instead the theory would be with 10 derivatives, in $d=6$, then addition of quintic killer terms would definitely solve the conditions for UV-finiteness. (This is analogous to the construction of 8-derivative QG UV-finite model in $d=4$ as presented in \cite{Modesto:2014lga,Modesto:2015lna,Modesto:2017sdr}.) Additionally, one can add some matter or (super-)symmetries to make the model of \cite{Knorr:2021lll} completely UV-finite and with the vanishing anomaly coefficient $b_6=0$. Such conclusions can be extended to any minimal renormalizable theory containing terms with up to $d=2n$ derivatives of the metric in $d$ spacetime dimensions. In particular, this explains the difference between full AS and only partial UV-finiteness of the Stelle quadratic gravity in $d=4$ as analyzed in \cite{Codello:2006in}.

As we emphasized above in the six-derivative gravitational model studied in this article, we will not have any logarithmic perturbative UV divergences proportional to the terms whose front coupling coefficients play the role of the Newton's gravitational constant or of the cosmological constant. Therefore, we will not discuss the findings related to the Reuter's FP of gravitational theory giving the first evidence for asymptotic safety scenario in gravity \cite{Reuter:1996cp}. Actually, we know that these results give a theoretical solution to the theoretical problems of non-renormalizability of the E-H action in $d=4$ spacetime dimensions \cite{Lauscher:2001ya}. However, we think that perhaps there is no any observable physics at intermediate energy scales behind them and they are only purely theoretical developments. First and foremost we believe that the true quantum gravitational theory should be with higher derivatives in their defining actions. The corresponding strong arguments given by DeWitt and Utiyama are already quite old \cite{Utiyama:1962sn}. Moreover, these results as obtained in \cite{Gies:2015tca} are scheme-dependent and also gauge- and gauge-fixing dependent, hence they cannot have any true gauge-invariant physical meaning. Moreover, they are read from ambiguous power-law like divergences in the perturbative calculus. There they are read for dimensionful couplings and this typically creates more problems with their universality in the framework of RG flows.

Instead of looking for something like the Reuter's FPs for dimensionless Newton's constant and dimensionless cosmological constant \cite{Dou:1997fg}, we would prefer to look for completely unambiguous situation of the RG flow of the couplings $g_R$ in front of the $R^2$ and $g_C$ in front of the $C^2$ terms, respectively. Their RG flows is gauge-independent, universal and unambiguous in six-derivative gravitational theory as we explained in the next section of this contribution. In this way we will look for analogous of FP as first found by Fradkin and Tseytlin for conformal supergravity four-dimensional quantum models \cite{Fradkin:1985am}.

Here, we remark that in a general situation the RG running is not physical and is not observable quantity. Additionally, the location of the FP is not physical, but the critical exponents are observable. However, first at the one-loop level the running of dimensionless coupling parameters is physical and observable and it has to do with the counting of perturbative degrees of freedom of the theory (both real and virtual in a specific sense). Moreover, the running at the one-loop perturbative level around the Gaussian FP is completely universal since any coupling redefinition, which preserves the perturbativity of couplings cannot change the results for these beta functions. In a more special situation it is possible to give an unambiguous physical meaning to the running of couplings, if the symmetry is involved. For example, the NSVZ exact beta functions are universal and observable since they have a very beautiful geometrical interpretation. As explained in \cite{shif1,shif2} they are related to the counting of zero modes, the dimension of the corresponding moduli spaces, and also they express the number of the instantonic zero modes \cite{Fradkin:1983qk}, which, in turn, is related to the number of nontrivially realized and hidden symmetries in the model. They have the geometric nature implying that the one-loop exact expressions for beta functions of K\"{a}hler sigma models have universal observable meaning despite that their expressions are fully non-perturbative. Being observable quantities, they are also renormalization-group~invariant.

The similar situation also happens in the case of six-derivative gravitational theories, where due to the enhanced renormalizability properties (so called super-renormalizability as explained in Section \ref{s2}), the exact expressions for beta functions, which happen to coincide with one-loop expressions, have also true physical meaning. There is a hidden symmetry behind this which lets everyone to give them a new geometrical interpretation. Only in such situation we can give a sensible meaning to the one-loop expressions for six-derivative gravitational theories and for the FPs of RG found there. Here, these quantities play the role of observables, because already in them one can show theoretically that all scheme dependencies mentioned above drop out, as it should be. Precisely in this situation, these beta functions, RG flows and the position of the FPs may have an invariant physical and geometrical meaning. The analogous conditions occur for the case of FPs of RG found first by Fradkin and Tseytlin in \cite{Fradkin:1985am} in the model of 4-dimensional conformal supergravity. There when one concentrates on only dimensionless coupling constant parameters (in front of the square of the Weyl tensor $C^2$), one derives the same universal conclusions about the RG flow, FP structure, critical exponents, etc.

One of the most disappointing features  of RG flows for made dimensionless Newton's constant $G_N$ (or cosmological constant $\lambda$) is their gauge dependence of special FP values~\cite{Falkenberg:1996bq}. Additionally, also the expressions for anomalous dimensions of various operators show some spurious dependence. These are some unacceptable statements in any gauge theory since they cannot simply lead to any definite observable physical consequences being gauge-variant (it does not matter whether this dependence is strong or only light, any dependence is wrong and incorrect). Such ideas to work with physical effects of gauge-dependent quantities clearly contradict from the definition the postulate of gauge invariance for relativistic field theories. After all, in such gauge models, one were able to treat gauge invariances in a relativistic manner but the price was to introduce a lot of gauge redundancy baggage. This was of course under the proviso, that this baggage will be completely unobservable and will not have any impact on physical effects of the theory. The relativistic formalism of QFT was rescued by the condition that everything gauge-dependent will never be observable. Or in more clear words in the formalism we have added some spurious degrees of freedom, only needed to save and marry Lorentz symmetry and gauge invariance. However, they do not carry with themselves any physically relevant information. Now, we think that deriving physical consequences (like scattering amplitudes) from explicitly gauge-dependent detailed description of the purported UV FPs is wrong since the latter have to be scheme-independent. This task can be then at best viewed as an incorrectly set exercise and this will not have any impact on physics at all. One of the way to solve this issue with explicit gauge dependence is to use some special formalism like presented in \cite{Lavrov:2012xz,Barra:2019rhz,Lavrov:2019agp}. This is why in this paper we will not comment any more on the Reuter type of FPs, which were the original basis for the AS program in QG research \cite{Reuter:1996cp}.

We add that it is natural to expect that the FP values of couplings in general are not physical, in particular, couplings can always be
redefined or non-linearly rescaled, and thus it is completely normal that FP values depend on the scheme. This also happens in
condensed matter physics. At the same time, these dependencies must disappear in physical observables (i.e., in
scattering amplitudes like computed in \cite{Draper:2020knh,scattering}) and thus there these dependencies are not an issue. The existence of FPs must be an universal statement (i.e., gauge- and scheme-independent), while their precise location in the parameter space is not. In correctly computed scattering amplitudes (as in \cite{Draper:2020knh}) there should not be seen any dependence on the gauge-fixing, scheme or the method of calculations. This is actually a good check for the formalism in which they are computed. In standard QFT, the amplitudes are obtained from gauge-dependent Green functions. For example, in the perturbative approach the independence and universality of the former ones is secured by famous Weinberg theorem about scattering on-shell amplitudes.

Eventually, we comment that in our theories (in particular in six-derivative QG) some values of the Newton's constant and the cosmological constant must be attained in the IR regime. This is likely due to the mechanism of dimensional transmutation, when the RG flow is run towards lower energies, and the model transforms into something else---into some IR description of gravitational physics and there it is not any more six-derivative QG theory. Of course, in such a regime the conditions for AS or for UV-finiteness do not hold any more and the RG flow has already taken place. The effective IR description can be probably accurate if one assumes that this is just E-H gravity plus some small corrections. We explain that the values of $G_N$ and of $\lambda$ that are reached in the IR regime must probably arise due to the phenomena of dimensional transmutation, since these are for the dimensionful gravitational couplings. In this way our theories resemble very much the theory of strong interactions---QCD, where in the UV regime they are well described by the Yang-Mills theories together with some matter species and characterized completely by only one dimensionless Yang-Mills coupling parameter $g_{\rm YM}$. While in the IR regime of the QCD, the famous $\Lambda_{\rm QCD}$  scale arises and this is a dimensionful quantity of mass dimension and the theoretical most accepted mechanism for its generation is this mentioned above dimensional transmutation. We think that the same mechanism is at work also for dimensionful gravitational constants ($G_N$ and $\lambda$) and for this the issue of an existence of UV FP is probably not very relevant. Instead, in this paper we concentrate on dimensionless couplings $g_R$ and $g_C$ and for them we do not need to invoke any special mechanism in the IR regime.

\section{Six-Derivative Gravitational Model}
\label{s2}

In this section we present briefly the main aspects of the considerations on the quantum level of the recently proposed model of six-derivative in the gravitational action in $d=4$ spacetime dimensions. For more details we refer the interested reader to the review paper in \cite{Rachwal:2021bgb}. Such six-derivative model of QG is a simple next order generalization of the original model as presented for Stelle in 1977. We consider it here on the quantum level since then it has some additional benefits in comparison to the general higher-derivative models of gravity. Roughly speaking regarding perturbative UV divergences this model behaves even better than the original Stelle gravity.

The Lagrangian of the six-derivative gravitational model reads
\begin{equation}
{\cal L} = \omega_C C_{\mu\nu\rho\sigma} \square C^{\mu\nu\rho\sigma} + \omega_R R\square R + {\cal L}_K.
\label{lagr}
\end{equation}

From this Lagrangian we construct the action of our higher-derivative QG model, here with six derivatives as the leading number of derivatives in the UV regime, by the formula\vspace{6pt}
\begin{equation}
S_{\rm HD} = \!\int\! d^4x\sqrt{|g|}{\cal L}.
\end{equation}

Above by $C_{\mu\nu\rho\sigma}$ we denote the Weyl tensor (constructed from the Riemann $R_{\mu\nu\rho\sigma}$, Ricci tensor $R_{\mu\nu}$ and Ricci scalar $R$ and with coefficients suitable for the case of $d=4$). Similarly we can write for the ``square'' scalar of the Weyl tensor
\begin{equation}
C^2=C_{\mu\nu\rho\sigma}^2=C_{\mu\nu\rho\sigma}C^{\mu\nu\rho\sigma}=R_{\mu\nu\rho\sigma}^2-2R_{\mu\nu}^2+\frac13R^2.
\end{equation}

Finally, to denote the box operator we use the symbol $\square$ with the following definition, i.e., $\square=g^{\mu\nu}\nabla_\mu\nabla_\nu$ which is a GR-covariant analogue of the d'Alembertian operator known from the flat spacetime case.

It is important to emphasize here that the  Lagrangian (\ref{lagr}) describes quite general
six-derivative theory describing the propagation of gravitational fluctuations around flat spacetime. For
this purpose it is important to include terms that are quadratic in gravitational curvature. As it
is obvious from the construction of the Lagrangian in (\ref{lagr}) for six-derivative model we have to
include terms which are quadratic in the Weyl tensor or Ricci scalar and they must contain precisely one power
of the covariant box operator $\square=g^{\mu\nu}\nabla_\mu\nabla_\nu$ (which is constructed using the
GR-covariant derivative $\nabla_\mu$). These two terms  exhaust all other possibilities  since all other terms
which are quadratic and contain two covariant derivatives can be reduced to the two above. This is achieved by exploiting
various symmetry properties of the curvature Riemann tensor as well as cyclicity and Bianchi identities.
The basis with Weyl tensors and Ricci scalars is the most convenient when one wants to study
the form of the propagator of graviton around flat spacetime. The QG model  with the Lagrangian (\ref{lagr}) is definitely the simplest one that describes the most general form of the graviton propagator around flat spacetime, in four spacetime dimensions and for the theory with six derivatives.

The special part of the Lagrangian in (\ref{lagr}) contain terms that are cubic in curvatures. The explicit form of them in $d=4$ after using all of symmetries and identities reads,
%\begin{adjustwidth}{-\extralength}{0cm}
%\centering %% If there is a figure in wide page, please release command \centering
\begin{eqnarray}
{\cal L}_K&=&s_1R^{3}+s_{2}RR_{\mu\nu}R^{\mu\nu}+s_{3}R_{\mu\nu}R^{\mu}{}_{\rho}R^{\nu\rho}
 +s_4RR_{\mu\nu\rho\sigma}R^{\mu\nu\rho\sigma}\nonumber\\
 &&+s_{5}R_{\mu\nu}R_{\rho\sigma}R^{\mu\rho\nu\sigma}+s_6R_{\mu\nu\rho\sigma}R^{\mu\nu}{}_{\kappa\lambda}R^{\rho\sigma\kappa\lambda}\,.
 \label{killers}
 \end{eqnarray}
%\end{adjustwidth}
 
 These terms are essential for killing the beta functions hence the common name for them---killers. As one can see they are all constructed as terms cubic in gravitational curvatures, while here six various contractions are possible. The coefficients $s_i$ will be determined in a moment from the condition of all-loop (or exact) UV-finiteness.

One can make the following additional remark here. Looking at the kinetic terms in the Lagrangian (\ref{lagr}), one can see that this type of construction bears great similarity to other higher-order derivative models, for example Ho\v{r}ava gravity (and Einstein--Aether theory to a lesser extent \cite{Jacobson:2010mx,Haghani}). These models were analyzed, for example, in \cite{Pinzul:2010ct,Lopes:2015bra,Pinzul:2016dwy}, where a special attention was put on their coupling to matter sectors. Moreover, some higher-derivative anisotropic operators were constructed and discussed in \cite{Mamiya:2013wqa}.

The six-derivative model, as presented here, is to some extent similar to the original higher-derivative model of Stelle quadratic gravity in $d=4$ \cite{Stelle:1976gc}. As emphasized before, higher derivatives are inevitably generated from matter quantum loops \cite{Utiyama:1962sn}. However, compared to the case of quadratic gravity, the model described by the Lagrangian in (\ref{lagr}), possesses also some significant differences due to the ``higher than minimal number of higher derivatives''. For renormalizable models in $d=4$, the dynamics with four derivatives is enough to achieve full renormalizability. By increasing the number of derivatives, the theory is still renormalizable. However, as we will describe in details below, the theory with six and more derivatives is super-renormalizable. Hence it possesses more distinct features of better convergence of loop integrals in the UV limits. Finally, this leads to the first UV-finite model of pure quantum gravity in $d=4$. As we emphasized in the introduction such a goal was not possible to be reached in the minimal model of quadratic gravity with four derivatives in $d=4$. In other words, we can term the theory from (\ref{lagr}) as described by ``higher than higher-derivative'' quantum gravitational model.

 Before presenting the final results for UV divergences of this theory, we briefly discuss the quantum properties of the theory based on the Lagrangian (\ref{lagr}). First, from the form of perturbative propagator around flat spacetime $\eta_{\mu\nu}$ for the graviton field, so for the fluctuations of the metric $h_{\mu\nu} = g_{\mu\nu} - \eta_{\mu\nu}$, we see that in the UV regime there is a suppression of the form $k^{-6}$. This is higher fall-off of the propagator than in standard Einsteinian theory  and also than in Stelle quadratic theory with four derivatives, where it is like $k^{-4}$. This already implies that the theory has better control over perturbative UV divergences.

 From the more precise analysis of power counting of UV divergences one sees a few important features here of the six-derivative model. First, this theory is renormalizable~\cite{Buchbinder:2021wzv} and the bound on the dimensionality of counterterms and of the superficial degree of divergences is $\Delta\leqslant4$. For a situation at the generic loop order $L$, the following bound on $\Delta$ is found: $\Delta\leqslant4-2(L-1)$ \cite{Rachwal:2021bgb}. This signifies that when higher loops are analyzed ($L>1$), then there is more and more restricted form of possible GR-covariant UV divergences of the model. In what follows below, we will only analyze for simplicity counterterms built out with terms containing precisely four derivatives of the background metric tensor. The other counterterms with smaller number of derivatives receive contributions also from higher loops (up to the order $L=4$ when the quantum theory is completely UV-finite).

 From the bound $\Delta\leqslant4$ we derive that UV divergences at any loop level $L$ can at most contain four derivatives of the metric. Since we read the beta functions of running couplings in the perturbative framework from corresponding UV divergences of the model, then we immediately conclude that couplings $\omega_C$ and $\omega_R$ in front of terms with six derivatives in the Lagrangian in (\ref{lagr}) never show any RG running behaviour. Hence these are constant non-running coupling parameters of the theory. Therefore, this implies that the situation with RG flow in this quantum model is simpler (than in Stelle theory for example) and is not so involved, despite that in the original definition of the model in \ref{lagr} we have significantly more couplings. In this way, we reduce the non-linearity of the system of equations for the RG evolution of couplings and one can solve it hierarchically starting with the couplings in front of terms with six derivatives (these are constants). Next are couplings for the terms with four derivatives and after solving for them we can read unambiguously the RG running of the effective Newton's constant $G_N$ (as described in \cite{Modesto:2017hzl}). Finally, knowing the RG flow for effective $G_N$ one derives and solves for the RG equation for the running cosmological constant $\Lambda$ (as described in \cite{Asorey:1996hz}).
 We remark that in the minimal super-renormalizable model as described by the Lagrangian \ref{lagr}, there are no subleading terms with four, two or zero derivatives on the metric and hence to the one-loop order in this model we do not have any unambiguous running of the couplings $G_N$ and $\Lambda$. Therefore, below we focus on UV divergences and related RG flows only for couplings $\theta_C$, $\theta_R$ and $\theta_{\rm GB}$ in front of terms in the divergent action with precisely four derivatives of the metric.

 When one considers only terms in the UV-divergent action with dimensionless couplings (for which the condition of AS has gauge-invariant and independent meanings), then the form of the one-loop divergent action reads, following \cite{Buchbinder:1992rb}
 \begin{equation}
  S_{\rm div} = -\frac{1}{\epsilon}\!\int\!d^4x\sqrt{|g|}\left(\beta_R R^2+\beta_C C^2+\beta_{\rm GB} {\rm GB}\right),
  \label{sdiv}
 \end{equation}
where by ${\rm GB}$ we denote the famous Gauss-Bonnet term in $d=4$ dimensions. In the above formula $\epsilon$ is a regulator (related to the one often used in DIMREG scheme) that when it is sent to zero the infinities are seen. Second advantage is that the theory is super-renormalizable meaning only first few loops (a finite number of them) gives divergent result. However, more importantly, when one analyzes from the power counting formula for $\Delta$ the number of derivatives necessary in counterterms, then for terms of dimension $4$ (like these in the action in (\ref{sdiv})) one finds that only one-loop quantum corrections to the divergences contribute here. This means that for the terms $R^2$, $C^2$ and for $\rm GB$ only contributions at the one-loop level must be taken. Hence by knowing the results at the first loop we know the effect of performing all loop resummations. The results are therefore one-loop exact. The quantum one-loop super-renormalizability that we meet in this model is encouraging us and we can also easily made the theory's control over perturbative UV divergences complete -- that is by playing with the additions to the Lagrangian in (\ref{killers}) we can satisfy the conditions $\beta_i$ for all $i$ (i.e., $i=R,C,{\rm GB}$). Moreover, we do not need to study here higher~loops!

 We finally remark that the whole notion of super-renormalizability (differently than just ordinary renormalizability) is inherently related to the perturbative development of the quantum model and perturbative loop integrals evaluated for parametrically small values of the coupling constants. Hence for it is almost practically impossible to find a non-perturbative meaning like the AS is.

Other advantages for studying the divergences with dimensionless couplings in front in $d=4$ dimensions, are that in this super-renormalizable model, the expressions for one-loop ultraviolet divergences are completely gauge-invariant, independent and also they do not depend on any gauge-fixing parameters, nor of parametrization ambiguities for definition of quantum fields (whether it is $h_{\mu\nu}$ or $h^{\mu\nu}$ or some other combination taken as the basic quantum variable) or details of the scheme chosen for renormalization or regularization of the quantum theory \cite{Kallosh:1978wt}. Finally these leading UV divergences containing four partial derivatives on the metric tensors are completely independent on some subleading terms in the number of partial derivatives that can be added to the action in~(\ref{lagr}). This is again a consequence of power counting and of arguments of the character of dimensional~analysis.

In this way such beta functions for dimensionless gravitational couplings are pieces of genuine observable quantities that can be defined in super-renormalizable models of QG. They are universal, unambiguous, independent of spurious parameters needed to define the gauge theory with local symmetries, and moreover they are exact, but still being computed at the one-loop level in perturbation calculus. They are clearly  very good candidates for the good physical observable in QG models. Therefore all these nice features gives us even more encouragement towards analyzing the structure of such physical quantities, to understand this based on some theoretical considerations and to use this QG model for the comparative description between AS and UV-finiteness.

All these above nice features of the six-derivative QG model makes it further worth studying as an example of an exact and
non-trivial RG flows in QG. Here we have exactness of one-loop expressions for running $\theta_R(t)$,
$\theta_C(t)$ and $\theta_{\rm GB}(t)$ coupling parameters together with
super-renormalizability. This exact character of UV divergences is one of the most powerful and beautiful features of the
super-renormalizable QG model here. Therefore this model gives us a very promising theoretical
laboratory for studying RG flows in general quantum gravitational theories understood in the quantum
field-theoretical framework.

The computation of the UV divergences in the model (\ref{lagr}) was done recently in \cite{Rachwal:2021bgb}.
The details of the methods used to obtain them were presented to some extent in this recent article. The method consists basically of using the Barvinsky-Vilkovisky trace
technology~\cite{Barvinsky:1985an} to compute functional traces of differential operators giving the expression for the
UV-divergent parts of the effective action at the one-loop level. The main results were obtained in
background field method and from UV divergences in \cite{Rachwal:2021bgb} we read the beta functions of
running gravitational couplings. The results for them in six-derivative gravitational theory in $d=4$
spacetime dimensions were the main achievements there.

\subsection{Killer Terms}
\label{s2.1}

Now, we discuss in greater details the role of the cubic killer terms in the action~(\ref{lagr}) contained collectively in ${\cal L}_K$. These terms again will come with front coefficients $s_i$ (for $i=1,\ldots,6$) of the highest energy dimensionality, equal to the dimensionality of the coefficients $\omega_C$  and $\omega_R$. Hence they could contribute to the leading four-derivative terms with UV divergences of the theory as written in (\ref{sdiv}). The general form of them was given in the list in (\ref{killers}).

\textls[-15]{Actually, these terms are essential for making the gravitational theory with six-derivative actions completely UV-finite. However, for renormalizability or  super-renormalizability} properties of the model in (\ref{lagr}), these terms are not necessary, e.g., they do not make impact on the renormalizability of the theory and therefore should be regarded as non-minimal ones. The set of terms in (\ref{killers}) is complete in $d=4$ for everything what regards terms cubic in gravitational curvatures. This non-trivial statement is due to various identities in $d=4$ as proven in \cite{Mistry:2020iex}.

These cubic terms are also sometimes called ``killers'' of the beta functions since  they  have profound effects on the form of the beta functions of all dimensionless gravitational couplings in the theory. This is very easy to explain. These terms are generally of the type $s{\cal R}^3$ and they are added to the original Lagrangian of six-derivative theories, where the main terms were of the type $\omega {\cal R}\square{\cal R}$ (important for flat spacetime graviton's propagator). It is well known that to get UV divergences at the one-loop level one has first to compute the second variational derivative operator (Hessian) $\hat H$ from the full action. The contributions from cubic killers to it will be of the form $s{\cal R}$ or higher in powers of curvatures. The second step is to derive the logarithm of the Hessian and subsequently the functional trace of such a matrix-valued differential operator (i.e., ${\rm Tr}\ln\hat H$). Next, when computing the trace of the functional logarithm one uses the known expansion of the logarithm in a Taylor series according to
\begin{equation}
\ln(1+z)=z-\frac12 z^2+\ldots.
\end{equation}

Hence one needs to take for sure up to the square of the contribution $s{\cal R}$ to the Hessian from the cubic killer term. The third power would be too much because of the following observation---we must remember that we are looking for terms of the general type ${\cal R}^2$  in the UV-divergent part of the effective action at the one-loop level. Hence the contribution to $S_{\rm div}$ of the cubic killer would produce the addition of the following general type $f(s){\cal R}^2$, where yet unknown functions $f(s)$ can be polynomials up to the second order in the coefficients $s_i$ of these killers. Now, requiring the total beta functions vanish (for complete UV-finiteness)  we need in full generality to solve the system of some quadratic equations in the coefficients $s_i$. The only potential problem for finding coefficients of the killers can be that some solutions of this system reveal to be complex numbers, rather than real. However, we need to require that all $s_i$ coefficients to be real for the definiteness of the classical action (for example in the Euclidean case of the signature of the metric). Therefore this issue requires a subtle and more detailed mathematical analysis, but the preliminary results based on \cite{Modesto:2011kw,Modesto:2014lga,Rachwal:2021bgb} show that in most of the cases the UV-finiteness is possible and easily can be achieved by adding some cubic killer operators from (\ref{killers}) with real coefficients $s_i$.

Below for definiteness we provide a sample result for these coefficients $s_i$ which make the full QG model from (\ref{lagr})  completely UV-finite. One can choose the following values for the coefficients $s_i$ and a relation $\omega_R=\omega_C$,
\begin{equation}
s_1=s_2=0,\quad s_3=5, \quad
 \frac{s_4}{\omega_C} \approx -0.847625, \quad \frac{s_5}{\omega_C} \approx 2.1177, \quad \frac{s_6}{\omega_C} \approx -9.83078.
 \label{ratios}
\end{equation}

This choice is not by any means unique and one should scan a space of parameters for other numerical solutions. The figures above are given also by exact formulas from solving some resulting cubic equations, and using Cardano formulas, but we will not need them here. One finds that real coefficients are quite generic solutions here to the system of three quadratic algebraic equations coming from requirements that $\beta_R=\beta_C=\beta_{\rm GB}=0$ in (\ref{sdiv}).

One can compare the situation here with cubic killers to the more known situation where the quartic killers are used to obtain UV-finiteness of gravitational theories. The second approach seems to be preferred one since the contribution of quartic killers (of the type schematically as ${\cal R}^4$) in $d=4$ to UV divergences is always linear  and proportional to ${\cal R}^2$ schematically. To solve a linear system of equations with linear coefficients, one always finds solutions and they are always real. This approach was successfully applied to QG theories in \cite{Modesto:2014lga}, to gauge theories in \cite{Modesto:2015foa}, to the theories on de Sitter and anti-de Sitter backgrounds \cite{Koshelev:2017ebj} and also in general non-local theories \cite{Modesto:2017sdr}. One shows that the UV-finiteness may be an universal feature of quantum field-theoretical interactions in nature \cite{Modesto:2015lna}. Moreover, this feature of the absence of perturbative UV divergences is related to the quantum conformality as advocated in \cite{Modesto:2016max,Modesto:2017xha}.

For swift comparison with other results we give also the expressions for perturbative one-loop UV divergences in $d=4$ for standard quadratic Stelle gravity \cite{Julve:1978xn,Fradkin:1981iu,Avramidi:1985ki}. {They are
shown below}
\begin{eqnarray} && \Gamma^{(1)}_{\rm div}
=
-\frac{1}{2\epsilon(4\pi)^{2}}\!\int\! d^4x \sqrt{|g|}
\left\{-
\frac{133}{20}C^2
+\left(-\frac{5}{2}x'^2+\frac{5}{2}x'-\frac{5}{36}\right) R^{2}+ \frac{196}{45}{\rm GB}
 \right\}\quad\quad\quad\\
 &&{\rm with}\quad x'=\frac{\theta_R}{\theta_C},\nonumber
 \label{stelledivs}
\end{eqnarray}
where in the original action of Stelle gravity the RG running couplings are $\theta_R$ and $\theta_C$ according to
\begin{equation}
 S_{\rm Stelle} = \!\int\! d^4x \sqrt{|g|} \left( \theta_R R^2+ \theta_C C^2\right).
\end{equation}

One sees that for this one-loop situation one can have without matter or additional symmetry arguments, only the restricted UV-finiteness in the $R^2$ sector of UV divergences. This is obtained for some special value of the ratio $x'=3(3\pm\sqrt{7})$ which approximate to {$x'_- \approx 1.06275$} or {$x'_+ \approx 16.9373$}. This is regarding the generic situation in quadratic four-derivative gravity. Such situation was also studied for the first time for the sake of conditions for AS in \cite{Codello:2006in}.

The authors of \cite{Codello:2006in}, in their paper indeed found AS for all couplings although these couplings were redefined and an overall rescaling of the total divergent action by the inverse Weyl coupling was performed. % to reach this conclusion. 
However, the point is the following. Even in the case of the model reaching UV AS as analyzed in \cite{Codello:2006in} the divergent part of the effective action was non-zero, or in other
words the coefficient of the trace anomaly, so called $b_4$ coefficient  in $d=4$ spacetime dimensions is non-vanishing.  We must admit
that the results of \cite{Codello:2006in} %authors of this PRL paper 
fully agree with the one-loop computation obtained
earlier by Fradkin, Tseytlin and independently by Avramidi in \cite{Fradkin:1981iu,Avramidi:1985ki}. 
The most clear interpretation of these results is that for none of the values of the $x'$ ratio, all independent terms in
(\ref{stelledivs}) (so $C^2$, $R^2$ and $\rm GB$ terms) can be cancelled out and hence this divergent part of the effective action is always
non-zero. Hence the pure quadratic theory is never fully UV-finite, and only in
the sector of $R^2$ divergences UV-finiteness is possible.

One would find it useful for pedagogical purposes
  to compare the situation in pure Yang--Mills
theory (YM) in $d=4$, where in the UV one meets AF, which is a subcase of AS. However, the one-loop
divergent effective action in this model is independent on the coupling and it is non-zero, hence the theory is not finite, although it reaches a (trivial) UV FP of RG flow. In such a model also the trace anomaly does not vanish ($b_4\neq0$) if it is coupled to the external metric. We think that the similar situation happens for the coupling in front of the Weyl tensor $C^2$ in Stelle gravity since its FP value also shows AF in this subsector of couplings. Therefore, in pure quadratic gravity, there is no conformal window possible for the model, while the AS in the UV regime is still an option. One of the important point of the present paper is that the vanishing of the full trace anomaly, divergent part of the effective action and consequently the existence of the conformal window for the pure gravitational model is achieved for the first time in the theory with six derivatives, which is analyzed here. One then can understand that this special UV-finite model sitting at the conformal window is a subcase of perturbatively realized AS models of quantum gravity.

In six-derivative gravity one has quite a similar situation but with some significant differences. For definiteness we consider here the situation completely without killer operators in (\ref{killers}). Then the running couplings are $\theta_R(t)$, $\theta_C(t)$ and $\theta_{\rm GB}(t)$, but they themselves do not source the RG flow since the last one comes only from non-running $\omega_R$  and $\omega_C$. In such a situation the exact one-loop results for UV divergences in 6-derivative gravitational theory in $d=4$ reads as follows\vspace{6pt}
\begin{equation} \Gamma^{}_{\rm div}
=
-\frac{1}{2\epsilon(4\pi)^{2}}\!\int\! d^4x \sqrt{|g|}
\left\{
\Big(\frac{2x}{9}+\frac{397}{40}\Big)C^2
- \frac{7}{36}R^{2}
+ \frac{1387}{180}{\rm GB}
 \right\},
 \end{equation}
 where in this case  the fundamental ratio of the theory is defined as $x=\frac{\omega_C}{\omega_R}$. In this case, only the $C^2$ sector of UV divergences can be made finite and this partial UV-finiteness is achieved for the value $x_*=-\frac{3573}{80} = -44.6625$ of the fundamental ratio. An attempt to discuss, understand and analyze the structure of these UV divergences (in comparison with divergences as presented in (\ref{stelledivs}) for Stelle gravity) an interested reader may find in \cite{Rachwal:2022hln}.

 In the  generic situation, the two couplings $\theta_R(t)$ and $\theta_{\rm GB}(t)$ run in such a way that in the deep UV regime asymptotic freedom is achieved in them. Their values tend for $t\to+\infty$ to infinity, respectively to $-\infty$ for $\theta_R$ and to $+\infty$ for $\theta_{\rm GB}$, so the initial conditions for the RG flows so we avoid Landau pole in  must be such that $\theta_R(t=0)<0$ and that $\theta_{\rm GB}(t=0)>0$. For the case of the $\theta_C(t)$ coupling the situation is more complicated. If the value of $x=x_*$, then this is an example of UV-finiteness and AS in this sector of $C^2$ divergences and this value of the non-running coupling $\theta_C$ is kept fixed at all energies, not only in the UV regime. We are then sitting at the non-trivial FP of RG. If non-running $x$ is smaller or bigger than  $x_*$, then there is a non-trivial running for $\theta_C(t)$ and the conditions for AF in the UV regime are similar to the stipulated above. If $x<x_*$, then $\beta_C<0$ and we must require an initial condition that $\theta_C(t=0)<0$, while in the opposite case if $x>x_*$, then respectively $\beta_C>0$ and the initial condition that $\theta_C(t=0)>0$ ensures that we meet AF in the deep UV regime without any dangerous Landau pole on our way towards it.

\section{UV-Finiteness Versus Asymptotic Safety in Six-Derivative Gravitational Model}
\label{s3}

Here we discuss and compare asymptotic safety approach to HD QG, which gives a lot of success for quantum field theory of
gravitational interactions treated both on the perturbative as well as on the non-perturbative level. We would like to compare this approach
with the other one offering full UV-finiteness of any Green function or scattering amplitude. This last case is realized by first making the
theory with six-derivative (so super-renormalizable) and later by adding some special operators cubic in gravitational curvature. One then
ends up with completely finite theories (without UV divergences). We mentioned already above that in the original Stelle theory we can
only make the $R^2$ sector of UV perturbative divergences at the one-loop level finite. Contrary, in the pure six-derivative theory (without terms cubic
in curvatures) it is possible by adjusting values of the fundamental ratio $x$ of the theory to make instead the $C^2$ sector of UV divergences completely finite.

In this section, we compare the conditions for UV-finiteness and for asymptotic safety based on the particular example of a six-derivative QG model as presented in the previous Section \ref{s2}. We also discuss the common practices that are used to extract some physical information from these two approaches towards the UV-completion of QG models. At first sight, there is a lot of similarities between these two research directions \cite{Percacci:2017fkn}. However, as we will point out also below there are some distinct differences, but this does not exclude the possibility for the unification and merging of the two approaches towards consistent QFT of quantum gravitational interactions.

First similarity is that in both approaches, one studies the same condition, namely
\begin{equation}
 \beta_i=0.
 \label{cond2}
\end{equation}

And this is the starting common point of two approaches. Here, we want to emphasize the differences. In models which are UV-finite, this condition of vanishing beta functions should hold for all running couplings of the theory and at all energy scales. This requirement may look as much stronger, but this is a true condition for the quantum model to look the same at any energy scale, hence there are no divergences possibly spoiling the conformal invariance of the quantum theory.

Instead, in the AS paradigm, the condition (\ref{cond2}) is to be put only in the UV regime of the RG flow. Only in the limits of infinite energies, we shall find a situation safe with UV divergences and only then we should reach a non-trivial FP, this is the asymptotic condition. Moreover, while in the approach with UV-finiteness, we require complete absence of all UV divergences, here for AS we may consider some sectors of couplings, in which there is a non-trivial FP, but some other couplings may run towards AF in the UV regime, and hence in the last case the UV divergences are still present (as also explained in full details below). We described previously that one can think of UV-finiteness as the special case of the AS, where the equation (\ref{cond2}) for all couplings must hold at all energy scales and not as a special autonomous condition in the parameter space. Therefore, the condition for vanishing of conformal anomaly are more special than just for the existence of the non-trivial FP of RG flow. Generically, the AS describes a special point in the parameter space, when the condition (\ref{cond2}) is solved for the running coupling parameters. (Actually at the FP they do not run anymore since they take their fixed valued there.)
In different vein by closely analyzing the perturbative asymptotic safety models as introduced by Litim and Sannino in \cite{Litim:2014uca}, one sees that the requirement there for perturbative AS is equivalent to the vanishing of UV divergences. Hence in such a situation the UV-finiteness for some sectors of couplings is a prerequisite for fully developed AS scenario. In the last perturbative model, where the AS is guaranteed, the UV-finiteness condition is put on the UV action since at lower energies the action may not be at the conformal window. It reaches the conformality only for asymptotically high energies due to the RG flow of dimensionless couplings. The consistency condition for the good behaviour in the infinite limit of energies (AS) is translated into the perturbative language as the condition for absence of ultraviolet divergences of perturbative loop integrals in the limiting theory described by the UV action. This also shows the complementarity between these two approaches to the problems of UV-completion and the UV limits of RG flows.
 
We remind here that in an example of AF theories, which are in a sense a trivial subset of AS theories, the perturbative UV divergences are present even in the UV regime of very high energies (the $b_4$ coefficient of the trace anomaly never vanishes since it is constant as computed to the one-loop level), while the perturbative couplings do run to zero values as it is required consistently with asymptotic freedom in the UV---the example of the QCD model is very paradigmatic here. To be more explicit when one writes schematically the YM action as $\alpha_F^{-1} F^2$, the perturbative coupling is $\alpha_F$, that is the loop expansion is in powers of this $\alpha_F$. One has at the tree-level the action proportional to the starting power $\alpha_F^{-1}$. Then at the first loop level, one finds that $\Gamma^{(1)}\sim\alpha_F^0$ -- is independent on the couplings and finally starting from two loops we see that for example $\Gamma^{(2)}\sim\alpha_F$ and in full generality $\Gamma^{(L)}\sim\alpha_F^{L-1}$. This confirms that again in pure YM theory the divergences are still there at the one-loop level and even in the situation when the trivial Gaussian FP is reached in the UV, and the coupling is zero, that is we have that when $k\to+\infty$ then $\alpha_F(k)\to0$, they are still present there and for example cause the non-vanishing of the trace anomaly. This is true even at the Gaussian FP where the couplings are formally zero. At the Gaussian FP we have that $\beta_{\alpha_F} =0 $, however for the inverse coupling $\omega_F=\alpha_F^{-1}$ needed to write the classical action in the form $\omega_F F^2$ there is no condition that $\beta_{\omega_F}=0$ at the FP. Instead, this last beta function is there constant ($\beta_{\omega_F}={\rm const}$), because it is constant at all energy scales to the one-loop accuracy.
 
 Therefore to refine our comparison between AS and UV-finiteness, we can say that they are very compatible to each other, except the AF sectors of the couplings of the theory, where the UV divergences generically still remain. In the last sectors the conditions for AF, despite the unambiguous presence of divergences as explained above, are still consistent with AS although they are not consistent with UV-finiteness. This is the only known  exception so far.

Moreover, we remark that when there is a FP (free or safe, or both) there
will be generically plenty of trajectories approaching this point. Only few special trajectories will reach the FP. Other trajectories will stay some RG time in the vicinity of the FP and eventually they will escape (they are not
UV-complete) and they correspond to inconsistent theories, for example whose UV divergence cannot be cured. If these
diverging trajectories did not exist (i.e., if the fixed point is fully UV attractive), the theory would have an
infinite (for infinite couplings in the truncation) or unbounded number of relevant directions  and thus it would not be predictive (and renormalizable), although it formally reaches a UV FP. Of course, this situation is not satisfactory.
On the diverging trajectories (not reaching the UV FP) the running of couplings under RG flow may lead to somewhere else (extreme regions) in the parameter space (like an asymptotic Landau pole for some couplings at infinite energies), where the theory is ill defined as well. Therefore also this situation should be definitely excluded from considerations of AS.

Therefore, here we would say that UV-finiteness is more restrictive condition because the FP is for all couplings and for all energy scales. One could even say that the theory without renormalization is easier to solve exactly, since what matters is mainly the classical level of such a model and the conformal symmetry helps with it.
An important essence of the AS program is that the emphasis should be put also on studies of the way how the FP is approached in the UV regime, not only the mere existence of the FP at the infinite energies. One could even risk a statement that UV-finiteness is needed for asymptotic safety at the close vicinity of the FP in order to have a full control over perturbative divergences in the scheme of expansion in small coupling parameters. The RG flow from the UV to the IR has also been studied in many papers in the framework of AS and in functional renormalization group (FRG), e.g., in
\cite{Gubitosi:2019ncx} (pure gravity and in the quadratic approximation) and also in \cite{Eichhorn:2019ybe}  (coupled gravity-matter systems). One can also explicitly integrate out the RG flow and compute the scale-dependent effective action $\Gamma_k$ as this was done for example in \cite{Codello:2015oqa,Ohta:2020bsc}. One of the goal of such a following of the flow towards the IR limit was to get the following limit $\Gamma_{k\to0}=\Gamma$, which gives the definition of the standard (quantum) effective action of the~theory. 

The second set of differences stems from the fact that as it was obvious from the example of quadratic Stelle gravity understood on the quantum level, it is impossible to have a fully UV-finite theory there, while in some sectors of couplings the AS is possible. At the end such a model of pure QG with only dimensionless coupling constants does not entail the possibility in $d=4$ to ameliorate the problem with divergences of QFT. In AS program, the models with dimensionless couplings in gravity are usually considered, however, as we pointed out on the example of six-derivative gravity from the previous section, the addition of higher order operators in gravitational curvatures is essential for the success of the construction of the UV-finite QG theories. To our knowledge the consistent addition of cubic or quartic operators in gravitational curvatures for the AS program in the framework of super-renormalizable QG models and the better control of the FPs in the UV regime was never studied before in the literature. Our papers (\cite{Rachwal:2021bgb,Rachwal:2022hln}) instead suggest that these terms are crucial for having better control of perturbative UV divergences. In QG models, which are super-renormalizable only divergences at the one-loop level matters and one possible way to kill them is to add these terms higher in curvatures. This is one solution in pure QG models, where the matter fields are not added.

Here one should also mention that there are studies of conditions for AS in models with additions of terms quartic and cubic  in curvatures (see e.g., \cite{Knorr:2021lll,Knorr:2021slg}), and also in the $f(R)$ approximation, where $R$ is the Ricci scalar of the spacetime and $f(x)$ is an arbitrary function. In the last truncation the beta functions  have been impressively computed up to the order $R^{70}$ \cite{Falls:2018ylp}. These studies add significantly to our understanding of situation with AS when cubic and quartic terms are included.
The inclusion of terms of the type $C \square C$ and $R \square R$ together with terms cubic in curvatures $O({\cal R}^3)$ was done here for the first time. We remind that the first group of terms (being quadratic in gravitational curvatures) have the impact on the improved
scaling of the propagator in the UV regime and it is entirely responsible for the super-renormalizability property of the
six-derivative gravitational theory. The addition of terms precisely cubic in gravitational curvatures
(and with no covariant derivatives acting between them) is motivated by effective field theory (EFT)
considerations since all such terms contain precisely 6 (partial) derivatives on the metric tensor after
expansion around flat spacetime and for the quantum field $h_{\mu\nu}$. In EFT all terms with given
number of derivatives must be potentially included for theoretical completion of the framework. The
expansion in number of derivatives is also counted by powers of the high energy scale $M$, so this is a
kind of derivative expansion or an expansion in the inverse powers of the big mass parameter $M$.
Finally, we remark that the terms with higher powers like $R^4$ or up to $R^{70}$ are not needed to
be explicitly added here, if one works to the order of six derivatives. Moreover, the basis of all terms as
presented in (\ref{killers}) is complete in $d=4$ spacetime dimensions since terms of all tensorial structure are present there
 (up to various identities holding in $d=4$). 

Another option is to add some matter fields (or supersymmetric multiplets of fields) in such a way that even pure QG model with dimensionless coupling parameters is UV-finite. The example here is mentioned before ${\cal N}=4$ conformal supergravity which is this highly supersymmetric analogue of scale-invariant Weyl gravity in $d=4$. In such a case without killer operators and without higher order terms, but in coupled QG model, one can find complete absence of UV divergences and presumably at the same time also conditions for AS in the UV regime. This model due to Fradkin and Tseytlin is probably one of the most promising for solving problems of QG and for having at the same time AS in the UV regime and UV-finiteness, together with local conformal symmetry and supersymmetry of matter and gravitational interactions. In general, however, classically scale-invariant QG models, may enjoy partial AS in some sector of couplings, but they are never fully UV-finite if the theory is in such a pure form. On this example one sees very well the differences between AS and UV-finiteness approaches to the problems of divergences in QFT models of QG.

One more difference can come from the observation that in AS scenario for QG, the FP values for the couplings are very special and they are grouped in isolated points on the parameter space (defined by coordinates being coupling parameters of the theory). Therefore the situation to find a FP is fine-tuned, and as the name suggests, these are points (so $0$-dimensional objects) in higher-dimensional parameter space of the theory. In contrary, the approach with UV-finite theory restricts couplings to lay on some one-dimensional curves being submanifolds in the bigger parameter space. Or they could be even higher dimensional submanifolds of FPs in super-renormalizable theories. From this point of view, one could say that the situation with UV-finite theory is less restrictive. However, another point of view is that having a fixed point or a line fixed points (or even a plane of them) does not make a difference, they are all solutions to the defining equation $\beta_i=0$. Instead, one could consider the situation in UV-finite theories as more fine-tuned since there the action of the model is restricted to have a few operators only, put there by hand, in order to have UV-finiteness. As we remarked previously, the UV-finite models can be viewed as special cases to the more broad class of UV-complete AS models.

The standard illustrative example one can recall here is given by the situation in ${\cal N}=4$  super-Yang--Mills theory, where if the coupling space is parametrized by Yang-Mills gauge coupling $g$, Yukawa coupling $y$ and the quartic self-interactions of scalars $\lambda$, then the ${\cal N}=4$  theory is obtained by putting two constraints on $g$, $y$ and $\lambda$. Such theory is fully conformal on the quantum level analyzed around flat spacetime background. There are then no UV divergences and the theory sits all the time at a FP. If the value of the coupling is non-zero, then this is a non-trivial FP. As one can see by counting constraints, the condition for UV-finiteness in this case is a one-dimensional curve, which can be parametrized for example by the value of the Yang--Mills coupling $g$. The other couplings are then functions of it.
The non-trivial FP is met for any value of the $g$ coupling, so this is strictly speaking a line of FP of RG and moving on this line does not cost any energy and there is no any RG flow within it.

A similar situation one also meets in the QG model with six derivatives. There the complete UV-finiteness (absence of the $R^2$, $C^2$ and the ${\rm GB}$ divergences) is obtained when the ratios $\frac{s}{\omega_C}$, as presented in (\ref{ratios}), are fixed, but the values of the coupling $\omega_C$ can be for this arbitrary, provided that this is not zero. We also find a line of FPs here and the condition for UV-finiteness does not determine uniquely the value of $s$, neither of $\omega_C$, only their ratio is fixed for the condition of UV-finiteness. Here we denote by $s$ a generic coupling in front of the killer operators from (\ref{killers}) admitting that in our case of six-derivative QG in $d=4$ there are six different non-running parameters $s_i$.

This last situation situation is reminiscent of what happens in AS paradigm to QG models, when one can rescale by the overall coupling the whole action. Then the couplings which appear there are basically the ratios and the condition for AS is insensitive to the overall coupling. If it was reintroduced there, then the condition for true AS would again define for us not just isolated points, but a line of them on the parameter space. However, we remark that for full UV-finiteness one cannot perform such procedure since if one rescales by the overall divergence of the action, then one loses all other divergences since relative coefficients between divergent terms of the effective action stay typically in finite mutual relations.

Finally, we must mention that for proper framing theoretically AS in the UV regime must embody the case of UV-finiteness. This is because of the presence of another issue within the AS program. This issue  which must be taken into account is that the presence of a non-trivial UV FP must be necessarily accompanied by the
UV-finiteness of the functional of the Wilsonian effective action when the last one is to  be computed at the formally infinite energy scales. Otherwise,
the whole construction of AS theory in the UV will not work and the UV divergences will still inevitably show up. Thus, AS would not be  UV-safe. This means that it must exist an UV action functional that is an UV endpoint of the RG trajectory as followed by Wilsonian effective action. One typically knows
that the UV effective action is a very complicated object since it may describe a theory with a lot of derivatives, or even a non-local one.
But just for its consistency such a functional must be UV-finite at the  infinite energy scale, in the opposite case the UV divergences are
still there.

The precise logical construction here is as follows. The UV limit of the Wilsonian effective action must exist as an action describing fully the UV physics, if this is understood as effective field theory. On the other hand, it must also exist as a well defined endpoint of the RG flow at energies $E_\infty$. Such action is typically call as the UV action. Sometimes people also call it as a bare or classical action of the theory, although in principle it has no relations to the classical action which was presumably there before the quantization process. In AS program it is often assumed that the UV action exists but its explicit construction is almost never showed. It is incorrect to think that it is the simplest action of the local theory in the UV limit. For consistent removal of divergences such local actions must come with very special properties. One can treat this UV action as a basis for quantum theory and ask the question of what happens in theoretically ``even higher energies'' $E>E_\infty$. For consistency of the whole  AS program, the UV divergences for ``even higher energies'' $E>E_\infty$ should be absent. (Here we assume the following sequence of limits, first $E\to+\infty$ and only later $E_\infty\to+\infty$).

Hence, we conclude that necessarily the functional for the UV action must be itself UV-finite when it is understood as the basis for some QFT. Otherwise we run back into a problem of pathological infinities.
Moreover, if one is able to provide an explicit construction for the UV-finite action, then one can give a satisfactory question to the old question in studies of RG flows---of what is the explicit form for the UV action for a consistent QFT which reaches a non-trivial UV FP. Previously, the existence of such actions was only hypothesized---now we can provide explicit examples of local UV action functionals, also in the gravitational frameworks.

Unfortunately, we remark that the issue of the proper construction of a true UV-finite UV Wilsonian effective action in the theories that reach AS in
the UV regime is typically neglected in the literature.  Maybe because one has not had good examples of UV-finite quantum gravitational theories.
They are not easily constructible using Einstein--Hilbert action or even Stelle gravity actions in $d=4$ spacetime dimensions. As we emphasized earlier for
this issue one must include \emph{higher derivatives} than four-derivative actions or add some very special matter content which probably needs to
be related also by some symmetry to the pure gravitational sector.

Complete UV-finiteness is not possible there without adding some special matter content (like in HD supergravity with ${\cal
N}=4$ due to Fradkin and Tseytlin). Instead the case of six-derivative theories is the first one in which one obtains for a model with
enhanced symmetry all beta functions vanishing, exactly, which here is equivalent to the situation at the one-loop due to
super-renormalizability properties. In this way one can find the example of a HD gravitational theory which completely describes the
situation at the UV fixed point of the RG flow. Hence one gives rise to a more detailed description of the AS program for QG by specifying an explicit example of the Wilsonian effective action that is there at the UV FP. One now understands that for the AS to work at the very high energies it is not enough that a hypothetical FP exists, the critical surface is finite dimensional, but also that there exists an action which describes the physics precisely at the UV FP and which is completely without UV divergences.

After all, we remark that the title of the seminal paper by Weinberg \cite{Weinberg:1980gg} was about UV divergences in gravitational theories. Of course, this paper started the whole revolution related to the AS program. The theoretical problem considered there was exactly with UV divergences and how to solve them to get the quantum theory UV-finite and UV-safe and as Weinberg coined it---asymptotically safe meaning safe from UV divergences (so finite) at asymptotically high energies. This is in close similarity to the origin of the terminology for AF theories, which are free for asymptotically high energies, so in the deep UV regime. We think that it is worth, following Weinberg, to come back, to unify and to appreciate again UV-finiteness considerations for asymptotically safe models of QFT since this is a necessary ingredient for the latter ones.

In this section, we discussed and compared the asymptotic safety approach to HD QG with the other one offering full UV-finiteness of any Green function or scattering amplitude. This last case was realized by first making the theory with six-derivative (so super-renormalizable) and later by adding some cubic killer operators to end up with completely finite theories (without UV divergences). We mentioned that in the original Stelle theory we can only make the $R^2$ sector of UV perturbative divergences at the one-loop level finite, and complete UV-finiteness is not possible there without adding some special matter content (like in HD supergravity with ${\cal N}=4$ due to Fradkin and Tseytlin). Instead the case of six-derivative theories is the first one in which one obtains for a model with enhanced symmetry all beta functions vanishing, exactly, which here is equivalent to the situation at the one-loop due to super-renormalizability properties. In this way one can find the example of a HD gravitational theory which completely describes the situation at the UV fixed point of the RG flow.

\section{Conclusions}
\label{s4}

In this paper, we described the two approaches towards a consistent formulations of quantum field theories of gravitational interactions valid at arbitrary high energy scales. We discussed both UV-finiteness and also Asymptotic Safety as apparently different ways of achieving UV-completion of gravity without necessity of invoking string theory or any other non-local extensions of standard physics. They are not rivals to each other. On the contrary, they possess a lot of common features and only superficial differences that upon closer inspection and analysis show the two different sides of the same coin. We showed that these differences can be understood and explained and that the mathematical basis for the two approaches is very similar. It is however a fact that typically UV-finiteness is considered for theories which are perturbative in coupling parameters, while AS is inherently rather a consequence of non-perturbative phenomena since FP values of couplings are not needed to be assumed to be small. This points to the problem that methodologies used to study these two approaches are usually different like perturbative vs. non-perturbative physics.

We still think that these two approaches can be unified and that they need each other. For example, we emphasized  above that although the non-perturbative formulation of AS does not assume the smallness of any coupling parameter, it must be consistent with the results obtained in the perturbative framework where this smallness is assumed and exploited. One particular and pertinent finding here is that the UV action functional, which describes the physics at the UV FP of RG flow, which stays at the core of AS paradigm, must itself entail the model UV-finite, when it is considered in standard perturbation theory. This implies that for consistency and for explicitness the AS program must contain the results already known from the perturbative approach to the UV-finite theories. Although as we remarked before UV-finiteness may also regard the non-perturbative divergences. This is why we think we can properly call the AS scenario not merely as non-perturbative renormalizability, but as a non-perturbative UV-finiteness in some theories. It is obvious that the full non-perturbative results (in AS) cannot go in contradiction with the perturbative results (from UV-finite theories). This last statement is true generally.

Let us summarize here the main ideas of this paper. The goal was exactly to compare the two approaches
 (UV-finiteness and AS) to perennial problems of quantum gravity with the conclusion that they are very similar
  and should not be viewed as rivals to each other. In this article, we wanted to point out similarities
   and the potential differences may be traced back only to the differences between perturbative
   and non-perturbative approaches to QFT. We think the honest
conclusion now is that the class of perturbatively
UV-finite theories is a subset of the theories which are generally asymptotically safe, both in
non-perturbative or as well as in perturbative framework. Simply UV-finiteness is a more restrictive
requirement than just AS since the former is related to the vanishing of trace anomaly coefficient $b_4$
at all energy scales, and not only to the situation at an ultraviolet FP of RG reached at infinite
energies.

In the course of our investigations, we first provided  examples where these relations between AS and UV-finite theories can be studied both perturbatively and non-perturbatively. We pointed to the scalar-gauge-Yukawa theories, in which perturbative analysis shows that they can be with guaranteed AS and also with perturbative UV-finiteness. It was more difficult to find similar examples in the case of QG models. For this one necessarily needs to study higher-derivative modified models of gravity. We remarked that in pure gravity models, in $d=4$ these are gravitational models due to Stelle with second powers of gravitational curvatures and with precisely four derivatives in the action, UV-finiteness in all sectors is impossible. This is in a close similarity to the case of pure non-Abelian gauge gauge theory in $d=4$, which is known that cannot be finite at the quantum level. In the last case one has to add charged matter, and for higher loop finiteness the matter must be of both fermionic and scalar types and matter interactions must consist of Yukawa scalar-fermion interactions as well as also of quartic scalar interactions. The examples of gauge UV-finite theories are here highly supersymmetric ${\cal N}=4$ super-Yang--Mills theories and also less symmetric scalar-gauge-fermion models of Litim and Sannino. The same considerations for gravity are way more complicated.

For the case of QG, in the quest for UV-finite theories one very (super-)symmetric option was to consider conformal supergravity with ${\cal N}=4$ as the analogue of the super-Yang--Mills theory. Besides this very symmetric option we decided to study pure gravitational models but with higher than usual number of higher derivatives. The considerations of six-derivative quantum gravitational models comes here with advantages that they were super-renormalizable and UV-finiteness was achieved to all loop orders. Our expressions for the beta functions were exact since there are no divergences at higher loop level, moreover they were exact, scheme and gauge parameter and gauge fixing independent. Based on such expressions after adding killers (cubic terms in gravitational curvatures) one could easily get an UV-finite theory at the one-loop perturbative level. However, this result we extended to all perturbative loops by the power of the super-renormalizability argument since divergences were expected and checked to be present only at the first loop level. On this example of some explicit UV-finite models of six-derivative QG we were able to discuss various differences and also similarities between the AS and UV-finiteness approach to the role of perturbative and non-perturbative UV divergences in the theory. This was the main purpose of studying here such an example of UV-finite pure six-derivative QG theory.

Here, we analyzed and constructed all gravitational terms in the action with up to six~derivatives on the metric. Of course, one can take models with more derivatives (with a finite number of them giving rise to local higher derivatives theories) or even fully non-local models (with a formally infinite number of derivatives) \cite{Knorr:2019atm,Modesto:2017sdr}. We just restricted to 6-derivative models since in this hierarchy these are the simplest and they are the first after Stelle quadratic theory in $d=4$ spacetime dimensions. We showed that what was not possible for quadratic gravity is possible for six-derivative model and then the $b_4$ coefficient vanishes, there is no trace anomaly on the quantum level and the theory sits at the conformal window. As one can see from the the formula (\ref{ratios}), there is also a significant freedom (as opposed to fine-tuning situations) to satisfy the conditions for UV-finiteness since for example the values of the coefficients of $\omega_C$ and $s_i$ are not determined separately, and only their ratio is specified for the conditions that all three beta functions of dimensionless couplings are zero here. We emphasized previously in the text that this is one example of UV-finite theory, but actually there are an infinite amount of them. The preliminary analysis shows that the allowed region for UV-finiteness spans a 3-dimensional compact region in a  7-dimensional parameter space of all couplings of the Lagrangian in (\ref{lagr}) (one coupling is removed here due to the overall rescaling of the action by $\omega_C\neq0$ coefficient, for example). However, the analytic description of this region is a difficult algebraic task. Hence one could say that the conditions for UV-finiteness are quite generic ones and these infinity of models mean that they are not fine-tuned. Similar advantages we also find in models, which are AS in the UV regime.

 In some standard matter theories, like QCD for example, just perturbative
renormalizability or AS is enough to have a well defined matter model, which is quantum consistent.
However, as we repeated many times above, in quantum gravity we may look for UV-finiteness because of the problem with
spacetime singularities of classical gravitational theory. Moreover, for renormalizable models one can
verify experimentally the RG running of couplings, while in the UV conformal gravitational phase such phenomena could
not be seen, even in theory. We think this is one of the distinguishing features of these models. Besides this, in mere renormalizable models, the couplings take different values than in the non-UV-finite models, so we expect that predictions for some observables
will be different. Additionally, there is a vast literature on the implications of conformal windows for
models of particle physics phenomenology (e.g., \cite{Lombardo:2014pda,Ryttov:2007sr,Orlando:2020yii} and also in \cite{Antipin:2013exa,Antipin:2017ebo}), so there are definitely clear observational signatures here. 

One could also pose a question of how to recover the standard two-derivative Einsteinian gravity (popularly known as the first example of a generally relativistic theory) from this six-derivative model, which is moreover selected to be UV-finite after conditions in (\ref{ratios}). The answer is very simple since one can freely add terms subleading with the number of derivatives to the Lagrangian in (\ref{lagr}). In particular, one can add there the Einstein--Hilbert term, Stelle quadratic terms $R^2$ and $C^2$ or even a cosmological constant term. One can supplement the conditions for UV-finiteness since then two new beta functions are there to cancel (of the cosmological constant and of the Newton's coupling) according to the results from \cite{Asorey:1996hz,Modesto:2017hzl}. In this new situation one again may find full finiteness in the UV regime. On the other hand, when one goes towards IR the terms with smaller number of derivatives become more and more important and effectively the action of the theory takes the form of Einstein--Hilbert action (possibly with the cosmological constant term). In this way the known IR gravitational physics is fully recovered here.

Instead, the quest for UV-finite theories is a completely UV question. Then in such regime the theory may be completely different than the theory, which remains in the IR regime. As we know and expect (based on \cite{Utiyama:1962sn}, for example) in the UV regime higher derivatives are inevitable and then the UV action necessarily differs from the two-derivative Einstein--Hilbert action, which is safely recovered in the IR regime of the theory. We already emphasized in the introduction to this paper that the quantum conformality in the UV is needed for the consistent resolution of spacetime singularities\footnote{ Another options for a consistent resolution of GR spacetime singularities are provided by some non-local models or when one adds  ${\cal R}^3$ operators (and higher) since they are sufficient to solve black hole singularities problem even in the presence of the Einstein--Hilbert part of the action (see, e.g.,  \cite{Holdom:2002xy,Knorr:2022kqp}).}, which is first done on the classical level of conformal Weyl $C^2$ gravity and later this is not destabilized by quantum corrections, if the conformal symmetry is fully realized in the UV regime. It is exactly that regime, where the theory should reach the description of point-like singularities which are so ubiquitous problems for classical Einsteinian general relativity theory. This could be viewed as one of the most pressing reasons why the classical gravity is not a complete theory and why it has to be quantized right away. Moreover, the theory in the UV regime must be changed and it cannot be a two-derivative gravitational theory anymore.
We find that that theory which solves the problem of short distance singularities and at the same time of UV divergences must incorporate conformal symmetry on the quantum level. Hence, one of the necessary conditions is that the trace anomaly completely vanishes $b_4=0$ and one cannot bypass this condition by any overall rescalings. This gives rise naturally also to UV-finite theories. The examples of conformal anomaly-free theories were already well studied in the past \cite{Fradkin:1983tg}; however, in this work we extended the discussion for pure six-derivative gravitational theories (without matter species), where such UV-finiteness was possible for the first time.

We comment that the conformal symmetry is probably the last one to be found in the UV regime of particle physics. This is because probing it in principle requires access to infinitely high energies, or at least arbitrarily high. In the UV conformal gravitational phase all scales should be practically identical, so this is practically the same situation as at the FP of RG flow. The condition for cancellation of conformal anomalies is very powerful since then it implies that the theory is completely without divergences. We can eventually quote the famous words of late Paul Dirac about QFT. He believed that in a fully consistent quantum theory of the world all infinities present in QFT (like in QED on which he worked extensively) should be absent. This we achieved in UV-finite models of quantum gravity.

We admit that the situation at the UV FP, so formally at infinite energies, provided that it exists and that some UV action at it is UV-finite, is not the end of the story with QG. This only solves the pending and long standing theoretical problem with infinities within QG models. Of course, the last one due to conformal enhanced symmetry at the FP is also related to the issue of resolution of spacetime singularities. However, these two theoretical problems and their successful solutions within the QFT framework do not address the phenomenological consequences of QG models at some finite energy scales, for example around the Planck energy scales ($M_{\rm Pl} = 10^{19}\,{\rm GeV}$). However, the philosophy that we have in mind is the following. After solving all theoretical problems in the UV for QG, we finally could have a consistent theoretical model for QG as a QFT and within this we should run down in energies and check for some observable predictions of the theory. Then, we can make relations of these QG models to QG phenomenology, physics of the SM of particle physics coupled to gravity, unifications of gravity with other interactions, cosmology, black hole physics, sub- and trans-Planckian gravitational physics.

For this ambitious goal one has to leave the UV FP, and start a deformation of the UV-finite UV action functional for the gravitational theory. The conformal symmetry and scale-invariance which were the situation at the FP must at lower energies be somehow broken (there are various ways of doing this, cf. \cite{Jizba:2020hre,Jizba:2014taa}) and the ensuing RG flow will start immediately. The model will not be anymore at infinite energies, will not be at the UV FP, the running effective Wilsonian action \cite{Codello:2008vh,Codello:2015oqa,Ohta:2020bsc}
will be different than the action of UV-finite theory, the couplings will start to run down with the energy and they will not be anymore equal to their FP values. The RG flow will be back non-trivial, but the conformal perturbation theory (CPT) will allow for its studies around the non-trivial UV FP. Following such RG flows within the framework of CPT or FRG one should be able to reach physical energy scales like the mentioned above physical Planck scale $M_{\rm Pl}$. For this one can use the methods of Wilsonian (RG-running) effective actions $\Gamma_k$ (as done for example in \cite{Codello:2015oqa,Ohta:2020bsc}). There, at lower energies, one will not be in the AS regime, and because of this, the UV AS scenario will not be at work since one will not be there in the formally deep UV regime. (If one looks at the face value at the Wilsonian effective action, then  the quantum theory based on it likely will not be UV-finite). However, the form of the interpolating (effective) action at these energies will have imprinted on itself the conditions that the theory satisfied in the deep UV regime (AS and UV-finiteness) and then this means that such effective theory at the energies $E\sim M_{\rm Pl}$ will be very special and unique. We also remark that the full quantum
effective action $\Gamma$ describes the theory across all energies. For large energies, the couplings or form-factors \cite{Knorr:2019atm} reach the AS/UV-finite regime, for small energies one should recover GR, and for intermediate
energies (e.g., inflationary scale or Planck scale) one can describe the crossover or a smooth transition between the two
phases of the theory. We emphasize that at intermediate energies the two theories described by $\Gamma_k$, there with $k\sim M_{\rm Pl}$, are still consistent, since their UV-completion is achieved independently in the UV regime and the corresponding theories  at higher energies are well defined. The predictions of such theories at intermediate gravitational energies for observable gravitational physics will be of our next great interest.

This  is the final important link of the theory to the reality and to the observational physics. We plan to work on this in the future since the theoretical framework for QFT of gravity  within the UV-finite or AS is consistent and established now. Moreover, these two approaches seem to be very promising in solving the theoretical problems of QG within well tested field-theoretical framework. We are also in possession of explicit formulations of QG models that satisfy the above conditions both perturbatively and non-perturbatively. The link to QG phenomenology  will create a possibility and a necessity for some final tests or verification of the theoretical ideas for quantum modified gravity that we advocated in this work.

\vspace{6pt}
%%%%%%%%%%%%%%%%%%%%%%%%%%%%%%%%%%%%%%%%%%

\funding{This research received no external funding.}

\acknowledgments{We would like to thank I.L.S., L.M. and A.P. for initial comments and encouragement about this work. L.R. thanks the Department of Physics of the Federal University of Juiz de Fora for kind hospitality and FAPEMIG for a technical support. We also would like to acknowledge critical suggestions for improvements of the manuscript from an anonymous referee in the Universe journal. We believe with them the part of the description of asymptotic safety in gravity is now much more clear and honest. Finally, we would like to express our gratitude to the organizers of the ALTECOSMOFUN'21 conference (Alternative Gravities and Fundamental Cosmology) for accepting our talk proposal and for creating a stimulating environment for scientific online discussions.}

\conflictsofinterest{The author declares no conflict of interest.}

%=====================================

%\begin{adjustwidth}{-\extralength}{0cm}
%\centering %% If there is a figure in wide page, please release command \centering

\reftitle{References}

%\end{adjustwidth}

\end{document}